\documentclass[pdflatex,sn-mathphys-num]{sn-jnl}

\usepackage{latexsym}
\usepackage{graphicx}%
\usepackage{subcaption}
\usepackage{multirow}%
\usepackage{amsmath,amssymb,amsfonts}%
\usepackage{amsthm}%
\usepackage{mathrsfs}%
\usepackage[title]{appendix}%
\usepackage{xcolor}%
\usepackage{textcomp}%
\usepackage{manyfoot}%
\usepackage{booktabs}%
\usepackage{algorithm}%
\usepackage{algorithmicx}%
\usepackage{algpseudocode}%
\usepackage{listings}%
\usepackage{xspace}
\usepackage{float}
\usepackage{setspace}
\usepackage{tabularx}
\usepackage{array}

\theoremstyle{thmstyleone}%
\theoremstyle{thmstyletwo}%

\theoremstyle{thmstylethree}%

\newcommand{\mn}{\text{PFedEG}\xspace}
\newcommand{\ms}{\text{PFedEG*}\xspace}
\newcommand{\mpl}{\text{PFedEG+}\xspace}

\begin{document}
\renewcommand{\arraystretch}{1.2}
\title[Article Title]{Personalized Federated Knowledge Graph Embedding with Client-Wise Relation Graph}


\author[1,3]{\fnm{Xiaoxiong} \sur{Zhang}}\email{zhan0552@e.ntu.edu.sg}

\author[2]{\fnm{Zhiwei} \sur{Zeng}}\email{zhiwei.zeng@ntu.edu.sg}

\author[1]{\fnm{Xin} \sur{Zhou}}\email{xin.zhou@ntu.edu.sg}

\author[3]{\fnm{Dusit} \sur{Niyato}}\email{dniyato@ntu.edu.sg}

\author*[2,3]{\fnm{ZhiQi} \sur{Shen}}\email{zqshen@ntu.edu.sg}

\affil[1]{\orgdiv{Joint NTU-Webank Research Institute on Fintech}}

\affil[2]{\orgdiv{Joint NTU-UBC Research Centre of Excellence in Active Living for the Elderly}}

\affil[3]{\orgdiv{College of Computing and Data Science}, \orgname{Nanyang Technological University}, \orgaddress{\street{50 Nanyang Ave}, \city{Singapore}, \postcode{639798}}}


\abstract{Federated Knowledge Graph Embedding (FKGE) has recently garnered considerable interest due to its capacity to extract expressive representations from distributed knowledge graphs, while concurrently safeguarding the privacy of individual clients.
Existing FKGE methods typically harness the arithmetic mean of entity embeddings from all clients as the global supplementary knowledge, and learn a replica of global consensus entities embeddings for each client.
However, these methods usually neglect the inherent semantic disparities among distinct clients. This oversight not only results in the globally shared complementary knowledge being inundated with too much noise when tailored to a specific client, but also instigates a discrepancy between local and global optimization objectives.
Consequently, the quality of the learned embeddings is compromised. To address this, we propose \textbf{P}ersonalized \textbf{Fed}erated knowledge graph \textbf{E}mbedding with client-wise relation \textbf{G}raph (\textbf{\mn}), a novel approach that employs a client-wise relation graph to learn personalized embeddings by discerning the semantic relevance of embeddings from other clients.
Specifically, \mn learns personalized supplementary knowledge for each client by amalgamating entity embedding from its neighboring clients based on their ``affinity'' on the client-wise relation graph. Each client then conducts personalized embedding learning based on its local triples and personalized supplementary knowledge. We conduct extensive experiments on four benchmark datasets to evaluate our method against state-of-the-art models and results demonstrate the superiority of our method.}




\maketitle

\section{Introduction}\label{introduction}
A knowledge graph (KG) represents real-world facts in the form of triples like (head entity, relation, tail entity). Knowledge graph embedding (KGE) is a technique that represents entities and relations within a KG in a continuous vector space \cite{shen2022comprehensive}. This technique is conducive to numerous downstream applications such as KG completion \cite{ma2019elpkg}, disease diagnosis \cite{abdelaziz2017large,sang2018sematyp}, recommender systems \cite{wang2022multi}, and question-answering systems \cite{hao-etal-2017-end}. With the promulgation of general data protection regulation (GDPR) \cite{regulation2016679}, KGs from multiple sources are no longer stored centrally on a single device as a complete KG.
Instead, they are stored in a more decentralized manner across multiple clients. These distributed multi-source KGs are formally referred to as federated knowledge graphs (FKG).

Federated knowledge graph embedding (FKGE) aims to collaboratively conduct knowledge graph embedding learning across multi-source KGs while preserving data privacy, based on federated learning \cite{briggs2020federated,fede,fedec,fedkg,shamsian2021personalized}. Through federated learning, the server can aggregate entity embeddings from all clients.
The aggregated information then serves as external knowledge for each client, thereby improving each client’s local embedding learning.

\begin{figure}[h]
  \centering
    \begin{minipage}{0.44\linewidth}
		\centering
		\includegraphics[width=0.95\linewidth]{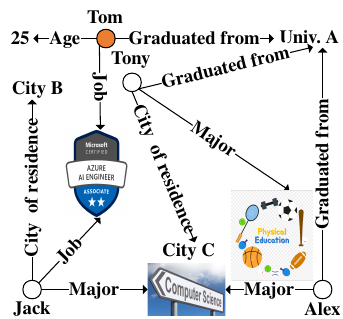}
		\subcaption{Company}
		\label{Company}
    \end{minipage}
     \begin{minipage}{0.48\linewidth}
		\centering
		\includegraphics[width=0.95\linewidth]{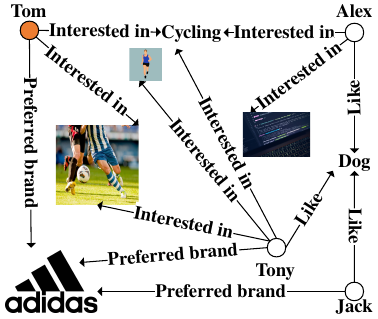}
		\subcaption{Club}
		\label{Club}
    \end{minipage}
    \begin{minipage}{0.8\linewidth}
		\centering
		\includegraphics[width=0.80\linewidth]{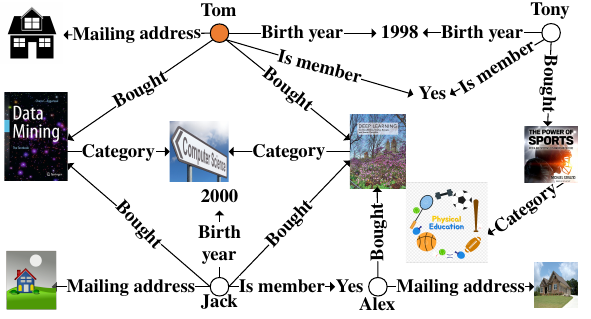}
		\subcaption{Book Store}
		\label{Book Store}
    \end{minipage}
  \caption{The illustration of a federated knowledge graph, where different KGs provide supplementary knowledge about entities (e.g. Tom).}
  \label{FKG}
\end{figure}

However, FKGE faces the critical challenge of semantic disparity among KGs, which can compromise the quality of the learned global embedding. This semantic disparity primarily derived from the variance in the relation set across clients. Distinct relations (or relation paths) represent diverse semantics for entities. As illustrated in Figure \ref{FKG}, the triple (Tom, Age, 25) in KG Company embodies ``age'' semantic for Tom and path $Tom \stackrel{Bought}{\longrightarrow} Deep~Learning \stackrel{Category}{\longrightarrow} Computer~Science$ in KG Book Store conveys ``major'' semantic for Tom. 
The relations instantiated in these three KGs do not exhibit overlap, and consequently, for a shared entity, its involved semantics may not be congruent across the three KGs. Existing FKGE methods typically fail to take this semantic disparity into consideration under the federated learning setting.

On one hand, prevailing methods usually employ an averaging strategy to amalgamate entity embeddings across all clients, thereby obtaining a replica of global supplementary knowledge that is universally shared among clients.
However, the resultant global supplementary knowledge may contain too much undesired information to certain clients. 
This is because, for a given client, the degree of semantic relevance of the embeddings derived from a diverse set of clients to this target client often exhibits significant  variation.
As depicted in Figure \ref{FKG}, the KG Company typically encapsulates entities’ occupational, residential, age-related, and academic data. 
In comparison to the KG Club which primarily focuses on entities’ hobbies, the KG Book Store appears to harbor more pertinent data to the Company as it implies age, residential and major information. This makes it more likely to supplement missing semantic data for entities within the KG Company.
Take entity ``Tom'' as an example, in the context of the KG Company, Tom lacks semantic data pertaining to his city of residence and major. However, Tom’s extensive history of professional book purchases within the KG Book Store provides an implication of his major. Similarly, Tom’s city of residence can be inferred from his mailing address within the KG Book Store. In terms of global supplementary knowledge about the entity ``Tom'', data procured from the KG Club is irrelevant to Tom within the context of the KG Company. Therefore, it is imperative that personalized supplementary knowledge is curated for each client through an examination of client-wise semantic relevance.

On the other hand, current methods typically learn a global consensus entities embeddings for all clients (Namely, for a shared entity $\textbf{e}$ across clients, different clients' entity $\textbf{e}$ learn the same embedding.), neglecting that the potential semantic disparitys among clients’ KGs can result in a divergence between local and global optimization objectives. The global entity embeddings are trained to fit the global KG, which is a composite of all clients’ KGs and encompasses all possible semantics. However, this approach may not generalize well on KGs that share only a few relations with the global KG, indicating a possible inconsistency between local and global optimization objectives \cite{zhang2023fedala}.
Consider the entity ``Tom'' in Figure \ref{FKG} as an example. The final learned embedding of Tom integrates semantic information about Tom from all three KGs. Despite the rich semantics of the global embedding about Tom, the semantic information about Tom's preference for sports (cycling, football, adidas) learned from the KG Club may lead the global embedding of Tom to inaccurately associate him more with sports-related majors, while in fact, he majors in computer science. Therefore, it is crucial for all KGs to collaboratively learn personalized or tailored embeddings.

In this paper, we propose a novel method named \textbf{P}er{sonalized} \textbf{Fed}erated Knowledge Graph \textbf{E}mbedding with Client-Wise Relation \textbf{G}raphs (\mn). Compared with existing FKGE methods, the server generates personalized supplementary knowledge for each client by aggregating entity embedding across clients based on their ``affinity'' within the client-wise relation graph. The ``affinity'' between two clients delineates the degree of semantic relevance between them, which is represented by the relation weight between these two clients on the client-wise relation graph.
This weight is learned through two proposed strategies, enabling a client to learn more from clients with higher ``affinity'' and less from those with lower ``affinity''. Furthermore, each client conducts personalized embedding learning with its personalized supplementary knowledge and local triples, ensuring that the learned embeddings are locally optimized while incorporating richer information from other clients.
We summarize the major contributions of this paper as follows:
\begin{itemize}
    \item We propose an approach, \mn, which aggregates entity embeddings across clients as personalized supplementary knowledge for each client based on the client-wise relation graph. \mn harnesses more relevant information from proximate Knowledge Graphs (KGs), thereby enhancing the quality of the learned embeddings for each KG.
    \item We emphasize the impact of the semantic disparity in FKG and are the first to propose conducting personalized embedding learning for individual KG leveraging personalized supplementary knowledge from other KGs, according to our best knowledge. 
    \item We evaluate the effectiveness of our method on four datasets by extensive experiments by comparing with existing methods. The results demonstrate a significant improvement in performance over state-of-the-art methods across four metrics that evaluate the accuracy of FKGE.
\end{itemize}

\section{RELATED WORK}

\subsection{Federated Knowledge Graph Embedding}
Federated learning is a distributed machine learning approach that allows multiple parties to collaboratively train a shared model while preserving data privacy and security \cite{fede,li2020federated,li2021survey,t2020personalized,sfl,ye2023personalized}. It has garnered attention in the research community for FKGE tasks. Current federated learning-based FKGE methods can be broadly classified into two categories: client-server architecture-based and peer-to-peer architecture-based. 

The client-server architecture involves the server coordinating local embedding learning for clients, while the clients engage in local training with local triples and aggregated entity embeddings on the server. FedE \cite{fede} is the pioneering model which adapts from the FedAvg \cite{mcmahan2017communication}.
The server aggregates entity embeddings from all clients in an average manner, and each client conducts local embedding learning based on local triples and the aggregated entity embeddings from the server. Inspired by Moon \cite{moon}, FedEC \cite{fedec}, based on FedE, introduces embedding-contrastive learning to guide clients' local training, thereby further enhancing the quality of learned embeddings.
However, both FedE and FedEC ignore the client-wise relation graph during the aggregation process of entity embeddings on the server, which has an adverse impact on the quality of learned embeddings. 

In contrast to FedE and FedEC, which primarily address the scenario where clients have partially shared entities but mutually exclusive relations, FedR \cite{fedr} specifically targets a different scenario where clients not only have shared entities but also shared relations. In this approach, all clients receive the same embeddings of shared relations from the server and subsequently conduct their local embedding learning using local triples and the shared relation embeddings.

In the peer-to-peer architecture, where there is no centralized coordinator like a server, clients collaborate on an equal footing and share embedding updates directly among themselves. To the best of our knowledge, FKGE \cite{fedkg} is the only model that operates in this manner and targets the same scenario as FedR. Drawing inspiration from MUSE \cite{lampleword}, FKGE employs a Generative Adversarial Network (GAN) structure \cite{goodfellow2020generative} to unify the embeddings of shared entities and relations within a KG pair. In this paper, we adopt client-server architecture rather than peer-to-peer architecture for communication efficiency.

\subsection{Knowledge Graph Embedding}

Knowledge graph embedding(KGE) is an approach to transform the relations and entities in an individual knowledge graph into a low dimensional continuous vector space \cite{choudhary2021survey, dai2020survey}, and use a score function to calculate the possibilities of candidate triples. Those candidate triples with high possibilities will be regarded to be true and used to complete KG. The motivation behind it is to preserve the structure information and underlying semantic information of the KG \cite{ma2019elpkg,rotate,ji2015knowledge,zhang2019iteratively,zhang2019interaction}. The learned embedding vectors by KGE models can be effectively applied in many downstream tasks, such as disease diagnosis \cite{sang2018sematyp,abdelaziz2017large}, question answering \cite{hao-etal-2017-end}, recommendation system\cite{zhang2016collaborative, xian2019reinforcement,wang2022multi} and knowledge graph completion \cite{bordes2013translating}. 

Existing KGE methods can be categorized into three categories: translational distance-based models \cite{bordes2013translating, transm,transr,rotate}, semantic matching-based models \cite{rescal,distmult,hole,complex} and neural network-based models\cite{rgcn,comgcn,convb}. The first category regards the relation in a triple as the translation between two entities. TransE \cite{bordes2013translating} and RotatE \cite{rotate} are representatives. TransE takes the relation as a vector translating the head entity to the tail entity for a true triple. The score function of the triple (h,r,t) is: $s(h,r,t) = -||\textbf{h}+ \textbf{r}-\textbf{t}||$ where $||\cdot||$ represents $L_1$ norm; $\textbf{h}$, $\textbf{r}$ and $\textbf{t}$ to represent vectors of head entity, relation and tail entity. RotatE \cite{rotate} defines the relation as a rotation from the head entity to the tail entity in the complex vector space. DistMult \cite{distmult} and ComplEx \cite{complex} are typical semantic matching-based methods. DistMult scores a triple by capturing pairwise interactions between the components of head entity and tail entity. ComplEx extends DistMult to model asymmetric relations by embedding relations and entities in a complex space. Recently, neural network-based models are also proposed to propagate neighborhood information so that generate more expressive entity and relation embeddings, such as graph neural network-based R-GCN \cite{rgcn} and convolutional neural networks-based ConvE \cite{convb}.

\section{METHODOLOGY}

In this section, we first describe the task of personalized federated knowledge graph embedding, and then present our method titled \textit{\textbf{P}ersonalized \textbf{Fed}rated Knowledge Graph \textbf{E}mbedding with Client-Wise Relation \textbf{G}raphs} (\textbf{\mn}) in detail.

\subsection{Problem Formulation}

Consider a federated knowledge graph, denoted as $\mathcal{G}_{Fed}=\{\mathcal{G}_c~| ~\mathcal{G}_c =\{\mathcal{E}_c, \mathcal{R}_c,\mathcal{T}_c\},$ $1 \leq c \leq C\} $, which is an aggregation of $C$ individual knowledge graphs distributed across $C$ clients. The three components in each $\mathcal{G}_c$ are denoted as $\mathcal{E}_c, \mathcal{R}_c$ and $\mathcal{T}_c$, which represent entity, relation and triple set of KG $\mathcal{G}_c$, respectively. 
These KGs are featured by partially overlapping entity sets. 

Personalized federated knowledge graph embedding (PFKGE) aims to learn personalized embeddings based on local triples and supplementary knowledge from other KGs without exposing raw triples explicitly to each other KG. Formally, the optimization objective for PFKGE is: 

\begin{align}
   \mathop {\min }\limits_{\{(\mathbf{E}_c,\mathbf{R}_c)|^{C}_{c=1}\}} \sum^{C}_{c=1} \left ( \frac{|\mathcal{T}_c|}{\sum^{C}_{i=1}|\mathcal{T}_i|} \mathcal{L}(\mathbf{E}_c, \mathbf{R}_c; \mathcal{G}_c,\mathbf{K}_c) \right ) 
\end{align}
where $\mathcal{L}$ is a loss function defined over a KG. $\mathbf{E}_c$ and $\mathbf{R}_c$ represent the learned personalized entity and relation embeddings for KG $\mathcal{G}_c$, respectively. $\mathbf{K}_c$ serves as supplementary knowledge, introducing external information to $\mathcal{G}_c$; $|\mathcal{T}_c|$ is the number of triples of KG $\mathcal{G}_c$.

In the absence of ambiguity, we use the two concepts of KG and client interchangeably in this paper, since each KG in $\mathcal{G}_{Fed}$ is stored on a client. Besides, in the following sections, if not specified, we use $C$, $N$, and $m$ represent the number of clients, the all unique entities from all clients and embedding dimension, respectively.
We assume that a private set intersection has given information about aligned entities~\cite{chen2020survey}, which is kept privately in the server. Besides, privacy is not the research focus since existing privacy-preserving methods (e.g. Differential privacy \cite{9069945}) can be incorporated into our methods.

\subsection{Overview of \mn}\label{overview}

Like existing FKGE methods, we also adopt the architecture of one server with multiple clients. Each client owns a KG and the server keeps the following information for coordinating clients: 
\begin{itemize}
\item An entity set $\mathcal{E}_{Fed}$, which is composed of unique entities from all clients. The size of $\mathcal{E}_{Fed}$ is $N$.

\item An existence matrix $\mathbf{M} \in \mathbb{R}^{C\times N}$. For each element $\textbf{M}_{ij}$, if $\mathbf{M}_{ij}=1$ or $0$, it means client $i$ has or has not entity $j$, respectively.  

\item A set of permutation matrices $\{\mathbf{P}^c \in \{0,1\}^{n_c \times N} ~|~ {1 \leq c \leq C}\}$, where $n_c$ is the unique entity number of client $c$. If element $\mathbf{P}^c_{ij} = 1(0)$, it means the entity $i$ of client $c$ (does not) corresponds to the  entity $j$ in $\mathcal{E}_{Fed}$. 
\end{itemize}    
\begin{figure}
\centerline{\includegraphics[height=2.5 in]{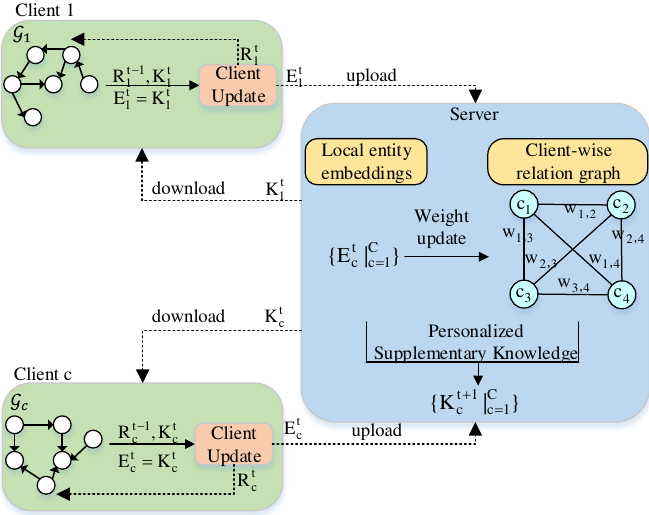}}
\caption{Overall procedure of one round in PFedEG. } \label{FedG}
\vspace{-1.8 em}
\end{figure}

The visual illustration and the pseudo-codes of PFedEG are shown in Figure \ref{FedG} and Algorithm \ref{alg:algorithm1}, respectively. 

There are mainly two phases: server update and client update, which are iteratively performed until the maximum number of rounds (line 2) or the performance of the learned embedding on the validation set does not increase for a certain number of rounds (line 9-11). At the beginning of each round, some clients are selected as a set $\mathcal{C}_{sel}$ with a certain selection fraction $\mathbf{F}$. During server update (line 14-19), the server is responsible for updating weights of the client-wise relation graph and aggregating personalized supplementary knowledge for each client using the local entity embeddings and the client-wise relation graph (line 15). During client updates, each selected client conducts personalized embedding learning by a KGE method based on its local triples and personalized supplementary knowledge from the server.

We will explain \mn in detail in the remaining parts of this section. 

\begin{algorithm}
\caption{PFedEG}\label{alg:algorithm1}
\begin{algorithmic}[1]
\Require $C$: number of clients, $\mathbf{F}$: client selection fraction, $E$: number of local epochs, $B$: local batch size, $\eta$: learning rate, $\mathbf{W}$: weight matrix, $\mathcal{T}_{m}$: maximal iteration rounds, $ESP$: early stopping patience.
\Ensure The learned embeddings for each client $c$.
\State Initialize embeddings $\mathbf{E}^0_c$ and $\mathbf{R}^0_c$ for each client $c$
\While{$round~t \leq \mathcal{T}_{m}$}
    \State $\mathcal{C}_{s} \Leftarrow$ (Set of randomly selected $\mathbf{F} \times C$ clients)
    \State $\{\mathbf{K}^t_c \,|\,^C_{c=1}\} \Leftarrow$ \Call{Server~Update}{$\mathbf{W}$, $\{\mathbf{E}^{t-1}_c \,|\,^C_{c=1}\}$}
    \ForAll{$c \in \mathcal{C}_{s}$}
        \State $\mathbf{E}^t_c, \mathbf{R}^t_c, \textit{MRR}^t_c \Leftarrow$ \Call{Client~Update}{$\mathbf{E}^{t-1}_c$, $\mathbf{R}^{t-1}_c$, $\mathbf{K}^t_c$}
    \EndFor
    \State $\textit{MRR}^t \Leftarrow$ WeightedAverage($\{\textit{MRR}^t_c \,|\, c \in \mathcal{C}_{s}\}$)
    \If{$\textit{MRR}^t$ drops for $ESP$ times in a row}
        \State \textbf{Break}
    \EndIf
\EndWhile
\State \Return $\mathbf{E}^T_c$ and $\mathbf{R}^T_c$ ($c \in [0, C]$, $T$ is the difference between last round and $ESP$)
\Statex

\Function{Server~Update}{$\{\mathbf{E}^{t-1}_c \,|\,^C_{c = 1}\}$}
    \State $\mathbf{W} \Leftarrow$ Relation weights update by Eq. \ref{ratio-based} (or \ref{emb-based}) \& \ref{scal}
    \State $\mathbf{G}^{t-1} \Leftarrow$ Transform $\{\mathbf{E}^{t-1}_c \,|\,^C_{c = 1}\}$
    \State $\{\mathbf{K}^t_c \,|\,^C_{c = 1}\} \Leftarrow$ Update personalized supplementary knowledge by Eq. \ref{aggregation},\ref{combine}-\ref{trans2}
    \State \Return $\{\mathbf{K}^t_c \,|\, c\in \mathcal{C}_{s}\}$
\EndFunction 
\Statex
\Function{ClientUpdate}{$\mathbf{E}^{t-1}_c$, $\mathbf{R}^{t-1}_c$, $\mathbf{K}^t_c$}
    \State $\mathbf{E}^t_c \Leftarrow \mathbf{K}^t_c$
    \State $\mathcal{B} \Leftarrow$ Split client triples $\mathcal{T}_c$ into batches of size $B$
    \For{$e \in [1, E]$}
        \ForAll{$b \in \mathcal{B}$}
            \State $\mathbf{E}^t_c \Leftarrow \mathbf{E}^t_c - \eta \Delta \mathcal{L}(b) \qquad \qquad \blacktriangleright$ Eq.\ref{loss}
            \State $\mathbf{R}^t_c \Leftarrow \mathbf{R}^t_c - \eta \Delta \mathcal{L}(b)$ (update relation weights)
        \EndFor
    \EndFor
    \State $\textit{MRR}^t_c \Leftarrow$ Evaluate $\mathbf{E}^t_c, \mathbf{R}^t_c$ on validation set
    \State \Return $\mathbf{E}^t_c$, $\mathbf{R}^t_c$, $\textit{MRR}^t_c$
    
\EndFunction

\end{algorithmic}
\end{algorithm}

\subsection{Client Update In \mn}

For each client, the client update aims to update its personalized entities and relations embedding based on its local triples, acquired personalized supplementary knowledge from the server and a knowledge graph embedding (KGE) method \cite{choudhary2021survey,dai2020survey,fan2014transition, transr,yang2014embedding}. In each round, all selected clients share the same client update process. Hence, we take the client $c$: $\mathcal{G}_c~=~\{\mathcal{E}_c,~\mathcal{R}_c,~\mathcal{T}_c\}$ at round $t$ as an example to explain the process.

Following \cite{rotate,fedec}, we adopt the KGE loss function with self-adversarial negative sampling to train each triple in client $c$. Besides, inspired by FedProx \cite{FedProx}, we not only initialize local entity embedding at the beginning of each training round with personalized supplementary knowledge as is calculated by the server in section \ref{su}, but also constrain the updated local entity embedding not far from the personalized supplementary knowledge with a regularization term, in order to make better use personalized supplementary knowledge. Compared with the regularization term, the initialization can directly leverage information of personalized supplementary knowledge, however the local entity embeddings may tend to forget external knowledge as local training. Hence, it is necessary to use both simultaneously. Formally, the training objective for client $c$ at round $t$ is as follows:
\begin{equation}\label{loss}
    \mathcal{L}(\mathbf{E}^t_c,\mathbf{R}^t_c; \mathcal{G}_c,\mathbf{K}^t_c) = \sum_{T \in \mathcal{T}_c} \mathcal{L}_{KGE}(T)~+~\beta D(\mathbf{E}^t_c,  \mathbf{K}^{t}_c),
\end{equation}
where $\beta \in \mathbb{R}$ is a hype-parameter; $\mathcal{T}_c$ is the set of raw triples and created triples by negative sampling; $\mathcal{L}_{KGE}$ is the KGE loss function with self-adversarial negative sampling; $D(\mathbf{E}^t_c,  \mathbf{K}^{t}_c)$ aims to constrain the update direction of local entity embedding and is defined as follows:
\begin{equation}\label{distance}
    D(\mathbf{E}^t_c,  \mathbf{K}^{t}_c) = \sqrt{\sum^{n_c}_{i = 1}\sum^m_{j = 1}({\mathbf{E}^t_{c}}_{(i,j)}-  {\mathbf{K}^{t}_{c}}_{(i,j)})^2},
\end{equation}
where $\mathbf{E}^t_c \in \mathbb{R}^{n_c\times m}$ and $\mathbf{K}^t_c \in \mathbb{R}^{n_c\times m}$ are client $c$'s local entity embeddings and personalized supplementary knowledge from the server at round $t$, respectively; $n_c$ is the entity number of client $c$; $(i,j)$ is two-dimensional index of $\mathbf{E}^t_c$ and $\mathbf{K}^t_c$.

\subsection{Server Update In \mn}\label{su}

The server update aims to aggregate entity embedding from all clients into personalized supplementary knowledge for each client by using relation weight in the client-wise relation graph. Each round of training has the same process and we take the round $t$ as an example to explain the process.   

The main challenge is how to quantify the 
``affinity'', i.e, the relation weight, between two KGs. Here, we put forward two strategies: (1) Shared entities number-based strategy; (2) Shared entities embeddings distance-based strategy.

For the former, the relation weight between two clients is defined as the ratio of the shared entity number to the total entity number of the two clients. For ease of calculation, we introduce a self-relation weight for each client, and it is assigned as the minimum value of the relation weight between this client and other clients. Formally, the relation weight $\mathbf{\hat{W}}_{i,j}$ between client $i$ and $j$ is :
\begin{equation}\label{ratio-based}
    \mathbf{\hat{W}}_{i,j} = 
    \begin{cases}
      \frac{|\mathcal{E}_i \cap \mathcal{E}_j|}{|\mathcal{E}_i \cup \mathcal{E}_j|},  & i \neq j;\\
      &\\
      \min(\{ \frac{|\mathcal{E}_i \cap \mathcal{E}_k|}{|\mathcal{E}_i \cup \mathcal{E}_k|}| k=1 \cdots C, \text{and}~k \neq j\}),  & i = j.
    \end{cases}
\end{equation}

For the latter, we take the cosine distance between shared entities embeddings of two clients as their weight. The weight is continuously optimized as the embedding update during iterative training. The  self-relation weight for each client is assigned as $e^{-1}$.  Formally, the relation weight $\hat{W}_{i,j}$ between client $i$ and $j$ is :
\begin{equation}\label{emb-based}
    \mathbf{\hat{W}}_{i,j} = 
    \begin{cases}
        \sum^{S}_{k=1} e^{cos( \mathbf{E}^t_{ij}(k), \mathbf{E}^t_{ji}(k))} &i \neq j;\\
        &\\
        e^{-1}&i = j,
    \end{cases}
\end{equation}
where $\mathbf{E}^t_{ij} (\mathbf{E}^t_{ji}) \in \mathbb{R}^{S \times m}$ is client $i$'s ($j$'s) shared entity embedding matrix with client $j$ ($i$) at round $t$; $S$ is the entity number of $\mathbf{E}^t_{ij}$; $\mathbf{E}^t_{ij}(k)$ is the embedding of $k_{th}$ entity in $\mathbf{E}^t_{ij}$.

For both strategies, the relation weight among clients are further scaled for the stability of the training. Take client $i$ as an example, the relation weight between client $i$ and $j$ is further scaled as follows:
\begin{equation}\label{scal}
    \mathbf{W}_{i,j} = \frac{\mathbf{\hat{W}}_{i,j}} { \sum^C_{j=1} \mathbf{\hat{W}}_{i,j}  }.
\end{equation}

Next, the server will aggregate entity embeddings from all clients as personalized supplementary knowledge for each client with the relation weights. Before that, the server first unifies entity embedding matrices from all clients into the same dimension by setting all zero vectors for nonexistent entities considering different clients usually has different entity sets. Formally, for each client's embedding matrix $\mathbf{E}^{t-1}_{c} \in \mathbb{R}^{n_c\times m}$ 
, it will be transformed as follows:
\begin{equation}
    \mathbf{\dot{E}}^{t-1}_{c} = {(\mathbf{P}^c})^\top \times \mathbf{E}^{t-1}_c,
\end{equation}
where $\mathbf{\dot{E}}^{t-1}_{c} \in \mathbb{R}^{N\times m}$; $\mathbf{P}^c$ is the permutation matrix as introduced in section \ref{overview}.

Then, the server concatenates $\mathbf{\dot{E}}^{t-1}_c$ from all clients into a two-dimensional global matrix $\mathbf{G}^{t-1} \in \mathbb{R}^{C\times(Nm)}$, by transforming each $\mathbf{\dot{E}}^{t-1}_c$ into a one-dimensional vector by rows and then stacking these vectors. 

With the global matrix and the relation weights, the personalized supplementary knowledge for all clients is calculated as follows:
\begin{equation} \label{aggregation}
    \mathbf{K}^{t} = Norm(\mathbf{W} \times \mathbf{G}^{t-1}, \mathbf{W} \times \mathbf{M}),
\end{equation}
where ``$\times$'' means matrix multiplication; $\mathbf{W}\in \mathbb{R}^{C\times C}$ is weight matrix and $\mathbf{M}\in \mathbb{R}^{C\times N}$ is the existence matrix as introduced in section \ref{overview}; $\mathbf{K}^{{t}}\in \mathbb{R}^{C\times(Nm)}$; For each entity $\mathbf{e}$, $Norm(\cdot)$ is used to normalize the sum of weights of clients owing $\mathbf{e}$ and is defined as follows:
\begin{equation}
    \mathbf{Z}=Norm(\mathbf{X}, \mathbf{Y}): \mathbf{Z}_{ij} = \frac{\mathbf{X}_{ij}}{\mathbf{Y}_{i,j//m}}.
\end{equation}
where ``//'' is floor division operator;  $\mathbf{Z}\in \mathbb{R}^{C\times Nm}$; $\mathbf{X}\in \mathbb{R}^{C\times Nm}$; $\mathbf{Y}\in \mathbb{R}^{C\times N}$. 

Inspired by ResNet \cite{he2016deep}, we further update personalized supplementary knowledge $\mathbf{K}^{t}$ by combining it with clients' local entity embeddings proportionally. Formally,
\begin{equation} \label{combine}
    \mathbf{K}^{t} = p \times \mathbf{K}^{t} + (1-p) \times \mathbf{G}^{t-1},
\end{equation}
where $p \in \mathbb{R}$.

The two-dimensional matrix $\mathbf{K}^{{t}}$ is the personalized supplementary knowledge for all clients, and the personalized supplementary knowledge for each client can be obtained by transforming $\mathbf{K}^{{t}}$ into $C$ matrices along the opposite direction to the transformation of $\mathbf{G}^{t-1}$. Formally, the element at index $(i,j)$ of personalized supplementary knowledge $\mathbf{K}^{t}_c\in \mathbb{R}^{N\times m}$ of client $c$ is as follows:
\begin{equation}\label{trans1}
\mathbf{K}^{t}_{c_{(i,j)}} = \mathbf{K}^{t}_{(c,i \times m+j)},
\end{equation}
where 
$(i,j)$ and $(c, i \times m +j)$ are two-dimensional indexes.

The $\mathbf{K}^{t}_c$ contains embedding of nonexistent entities due to setting all zero vectors for nonexistent entities for ease of calculation before, which should be further cleared. Formally, the final $\mathbf{K}^{t}_c$ is computed as: 
\begin{equation}\label{trans2}
    \mathbf{K}^{t}_c = {\mathbf{P}^c} \times \mathbf{K}^{t}_c,
\end{equation}
where ${\mathbf{P}^c}$ is the permutation matrix introduced in section \ref{overview}.

\section{Experiment}

\begin{table}[h]
\caption{Experiment results on FB15k-237-Fed10 and FB15k-237-Fed5. The \textbf{bold} figures denote the best results among methods except Collective.}\label{f35}
\setlength{\tabcolsep}{1mm}{
\begin{tabular*}{\textwidth}{@{\extracolsep\fill}llllllllllll}
\toprule%

\multirow{2}{*}{KGE} &  \multirow{2}{*}{Methods}  && \multicolumn{4}{c}{FB15k-237-Fed10} &&\multicolumn{4}{c}{FB15k-237-Fed5}\\
\cmidrule{4-7}
\cmidrule{9-12}

&&&MRR& Hits@1 & Hits@5 & Hits@10 && MRR& Hits@1 & Hits@5 & Hits@10 \\
\midrule

\multirow{6}{*}{TransE} & Collective&  &0.4111 &0.2764 & 0.5753  & 0.6719      && 
 0.4257 & 0.2964& 0.5815 & 0.6733  \\
  \cmidrule{2-12}
                        & Single&  & 0.3699& 0.2469 & 0.5164 &0.6066   && 
 0.3819 &0.2554 & 0.5343 & 0.6254 \\
                        & FedE&  &0.4050 & 0.2726 & 0.5652 & 0.6603  && 
  0.4163& 0.2892& 0.5671&  0.6598\\
                        & FedEC&  &0.4061 &0.2737  &0.5651  &0.6629   && 
  0.4206&0.2929 &0.5748 &0.6661\\
                        & \ms&  & $\mathbf{0.4228}$ & $\mathbf{0.2877}$ &0.5887  &$\mathbf{0.6838}$   && 
  $\mathbf{0.4294}$ & $\mathbf{0.2967}$ & $\mathbf{0.5906}$ & $\mathbf{0.6824}$ \\
                        & \mpl&  &0.4203 & 0.2841 & $\mathbf{0.5888}$ & 0.6830   && 
  0.4273& 0.2947& 0.5865&  0.6808\\
                       
\midrule

\multirow{6}{*}{RotatE} & Collective&  &0.4250 & 0.2899& 0.5901 & 0.6851     && 
  0.4414& 0.3121 & 0.5971 &  0.6864\\
  \cmidrule{2-12}
                        & Single&  &0.4026 & 0.2929 & 0.5326 &0.6159   && 
  0.4139&0.3011 & 0.5497& 0.6328 \\
                        & FedE&  & 0.4218&  0.2879&  0.5852&   0.6775&& 
  0.4320& 0.3024& 0.5881&  0.6785\\
                        & FedEC&  &0.4271 &0.2940  &0.5910  & 0.6834  && 
  0.4381& 0.3078& 0.5973& 0.6871 \\
                        & \ms&  & 0.4461 & 0.3142 & $\mathbf{0.6092}$ & 0.6985  && 
  $\mathbf{0.4500}$& $\mathbf{0.3241}$ &0.6031 &0.6934\\
                        & \mpl&  & $\mathbf{0.4473}$ & $\mathbf{0.3157}$ & 0.6091 & $\mathbf{0.6995}$  && 
  0.4495& 0.3216& $\mathbf{0.6054}$ &$\mathbf{0.6939}$\\
\midrule

\multirow{6}{*}{ComplEx} & Collective&  &0.4011 &0.2880 & 0.5355 &0.6253      && 
 0.4113 & 0.3043 & 0.5353 & 0.6255\\
 \cmidrule{2-12}
                        & Single&  & 0.3669 & 0.2655 & 0.4833 & 0.5674  && 
  0.3804& 0.2785& 0.4977&  0.5833\\
                        & FedE&  &0.3508 &0.2460  &0.4704  &0.5574   && 
  0.3752& 0.2722&0.4923 &0.5813  \\
                        & FedEC&  &0.3678 & 0.2581 & 0.4948 & 0.5871  && 
  0.3872& 0.2816& 0.5093&  0.5970\\
                        & \ms&  & 0.3779& 0.2702 & 0.5008 & 0.5913  && 
  0.3914&0.2871 &0.5118 &0.5981 \\
                        & \mpl&  & $\mathbf{0.3803}$ & $\mathbf{0.2723}$ & $\mathbf{0.5034}$ & $\mathbf{0.5940}$  && 
  $\mathbf{0.3961}$& $\mathbf{0.2908}$& $\mathbf{0.5169}$&$\mathbf{0.6042}$\\
\botrule
\end{tabular*}
}
\end{table}
In this section, we evaluate the performance of our method on real-world datasets through extensive experiments. We first describe the datasets. Then we introduce the benchmarks for our experiments and the corresponding experimental settings. Next, we quantitatively compare the effectiveness of our model with other state-of-the-art methods. Finally, we conduct ablation study and also show how parameter variations influence our method. 

\subsection{Data Description}\label{dd}
We conduct experiments with four datasets(FB15k-237-fed3, FB15k-237-fed5, FB15k-237-fed10 and NELL-995-Fed3), which are used for federated knowledge graph Embedding task and proposed by \cite{fedec}. For all clients in these four datasets, they follow the same training/validation/testing triples division ratio: 0.8/0.1/0.1. 

\subsection{Experimental Setup}\label{es}

\noindent \textit{\textbf{Baselines}}. 
\textbf{FedE} \cite{fede} is the first FKGE method, which averagely aggregates entity embeddings from all clients and each client conducts local embedding training with a KGE method based on local triples and the aggregated result. \textbf{FedEC}\cite{fedec}, based on \textbf{FedE}, introduces contrastive learning to local embedding learning. Besides, we compare our model with two settings: $\textbf{Single}$ and $\textbf{Collective}$. In the setting of $\textbf{Single}$, we conduct KGE for each client only with its local triples. In the setting of $\textbf{Collective}$, we collect triples from all clients and construct a new dataset. Although the setting of $\textbf{Collective}$ violates the constraint of data privacy, it can better reflect the ability of our model by comparing with it. We compare our method with previous methods based on three typical knowledge graph embedding (KGE) methods: TransE \cite{bordes2013translating}, RotatE \cite{rotate} and ComplEx \cite{complex}, which are often used for FKGE tasks. It should be noted that our method is not limited to these three KGE methods but are applicable to many others.

\begin{table}[h]
\caption{Experiment results on FB15k-237-Fed3 and NELL-995-Fed3. The \textbf{bold} figures denote the best results among methods except Collective.}\label{n3}
\setlength{\tabcolsep}{1mm}{
\begin{tabular*}{\textwidth}{@{\extracolsep\fill}llllllllllll}
\toprule%
\multirow{2}{*}{KGE} &  \multirow{2}{*}{Methods} && \multicolumn{4}{c}{FB15k-237-Fed3} && \multicolumn{4}{c}{NELL-995-Fed3}\\
\cmidrule{4-7}
\cmidrule{9-12}
&&&MRR& Hits@1 & Hits@5 & Hits@10 &&MRR& Hits@1 & Hits@5 & Hits@10\\
\midrule

\multirow{6}{*}{TransE} & Collective&  & 0.4334& 0.3063 &0.5859 & 0.6772  & &0.3628 &0.2231 & 0.5254 & 0.6271  \\
\cmidrule{2-12}
                        & Single& &  0.4070&0.2848&0.5543&0.6401 &  &0.3321 & 0.1949 & 0.4884 & 0.5865 \\
                        & FedE& &  0.4297&0.3036&0.5817&0.6692 &  &0.3489 & 0.2113 & 0.5075 & 0.6159 \\
                        & FedEC & & 0.4315 &0.3045&0.5866&0.6739 &  &0.3549 &0.2183  &0.5129  &0.6192  \\
                        &PFedEG* & & $\mathbf{0.4362}$  &$\mathbf{0.3077}$&0.5916&0.6815 &  &$\mathbf{0.3692}$ & $\mathbf{0.2307}$ & 0.5282 & $\mathbf{0.6337}$  \\
                        &PFedEG+& &  0.4357&0.3061&$\mathbf{0.5918}$&$\mathbf{0.6822}$ &  &0.3687 &0.2276  &$\mathbf{0.5337}$  &0.6331  \\
                       
\midrule

\multirow{6}{*}{RotatE} & Collective& &  0.4443 & 0.3198 & 0.5964& 0.6855  &  &0.4331 &0.3184 &0.5624  &0.6566  \\
\cmidrule{2-12}
                        & Single & &  0.4270&0.3106&0.5651&0.6489 &  &0.3858 &0.2804  &0.5031  & 0.5937 \\
                        & FedE & &  0.4427&0.3178&0.5957&0.6819 &  &0.4293 & 0.3072 &0.5705  &0.6630  \\
                        & FedEC& & 0.4480 &0.3226&0.6018&0.6891 &  &0.4354 &0.3184  &0.5702  &0.6631  \\
                        & PFedEG* & &  $\mathbf{0.4557}$&$\mathbf{0.3305}$&$\mathbf{0.6075}$&0.6937 &  &0.4387 &0.3247  &0.5669  & 0.6626 \\
                        & PFedEG+ & &  0.4539&0.3276&0.6070&$\mathbf{0.6963}$ &  &$\mathbf{0.4423}$ & $\mathbf{0.3285}$ & $\mathbf{0.5725}$ & $\mathbf{0.6675}$ \\
                       
\cmidrule{2-12}

\multirow{6}{*}{ComplEx} & Collective & & 0.4171 & 0.3104 &0.5413 & 0.6267 &  & 0.3729 & 0.2736 & 0.4827 & 0.5726  \\
\cmidrule{2-12}
                        & Single & &  0.3927&0.2923&0.5102&0.5919 & &0.3224  &0.2265 & 0.4313 & 0.5173  \\
                        & FedE & & 0.3907 &0.2893&0.5077&0.5897 &  &0.3493 &0.2478  &0.4649  & 0.5497 \\
                        & FedEC & &  0.4002&0.2976&0.5183&0.6030 &  &0.3667 &0.2634  &0.4844  &0.5712  \\
                        & PFedEG* & &  0.4050&0.2994&0.5298&0.6129 &  &0.3702 & 0.2692 & 0.4844 & 0.5687 \\
                        & PFedEG+ & &  $\mathbf{0.4086}$&$\mathbf{0.3017}$&$\mathbf{0.5355}$&$\mathbf{0.6197}$ &  &$\mathbf{0.3737}$ & $\mathbf{0.2725}$ & $\mathbf{0.4918}$ & $\mathbf{0.5721}$ \\
                       
\midrule
\end{tabular*}
}
\end{table}

In our experiments, we focus on predicting the tail entity when provided with the head entity and relation, namely $(h,r,?)$. All models are evaluated in terms of two representative metrics: Mean Reciprocal Rank(MRR) and Hits@N (N includes 1,5,10). With weight being the proportion of client triples, we use the weighted average of all clients' metric values as the final metric value. The default parameter settings across all experiments are as follows. For KGE loss function with the self-adversarial negative sampling (namely, $\mathcal{L}_{KGE}$), the temperature for negative sampling, the fixed margin and sampling size are 1, 10 and 256 respectively. The client selection fraction $\mathbf{F}$ is 1. For local update, the local epoch is 3 and the batch size is 512. The embedding dimension is 128. When training, the early stopping patience is 5 which represents the training will be terminated after 5 consecutive drops on MRR. For setting $\textbf{Single}$ and $\textbf{Collective}$, the performance of learned personalized embeddings is evaluated on validation sets per 10 rounds, whereas 5 epochs on the other models. With learning rates being 0.001, Adam \cite{adam} is used as the optimizer during client update. For dataset FB15k-237-Fed10, FB15k-237-Fed5, FB15k-237-Fed3, when KGE method is TransE or RotatE, the regularization coefficient $\beta$ is set as $3\times 10^{-3}$. For the other cases, the regularization coefficient $\beta$ is chosen from the set $\{3\times 10^{-4}, 4\times 10^{-4}, 5 \times 10^{-4}, 6 \times 10^{-4}, 1\times 10^{-3}\}$. The combination coefficient $p$ is chosen from the set $\{0.5,0.6,0.7,0.8,0.9\}$. The experiment code is available at \url{https://anonymous.4open.science/r/PFedEG-2BF9}.  

For clarity, we name the model using the shared entities number-based weight update strategy and the shared entities embeddings distance-based weight update strategy as PFedEG* and PFedEG+, respectively.

\subsection{Quantitative Analysis}

In this part, we quantitatively evaluate the performance of our method on the link prediction task, by comparing with the other baselines. The results on four datasets are reported in Table \ref{f35} and \ref{n3}. 

Through a comparative evaluation of performances of \ms and \mpl on four metrics against the best results of Single, FedE and FedEC, we find that both \ms and \mpl achieves better performance.   Specifically, when using TransE, RotatE, and ComplEx as KGE methods, the best result of \ms and \mpl achieves a relative rise in MRR on FB15k-237-Fed10 by 4.11\%, 4.73\% and 3.40\%, respectively. The corresponding relative rises in MRR on FB15k-237-Fed5 are 2.09\%, 2.72\% and 2.30\%, respectively. The dataset FB15k-237-Fed3 witnesses the corresponding relative rises in MRR by 1.10\%, 1.72\% and 2.10\%. The corresponding relative rises in MRR on NELL-995-Fed3 are 4.03\%, 1.58\% and 1.91\%, respectively. Analogous patterns of growth are also noticeable in Hits@1, Hits@5, and Hits@10. All rising figures demonstrate the potential of our proposed method to conduct FKGE task.

Besides, we also find that both \ms and \mpl even outperforms the results of Collective on the four datasets, with TransE and RotatE as KGE methods. Specifically, when using TransE and RotatE as KGE methods, the best result of \ms and \mpl achieves a relative rise in MRR on FB15k-237-Fed10 by 2.85\% and 5.25\%, respectively. The corresponding relative rises in MRR on FB15k-237-Fed5 are 0.87\% and 1.95\%, respectively. The dataset FB15k-237-Fed3 also witnesses the corresponding relative rises in MRR by 0.65\% and 2.57\%. The corresponding relative rises in MRR on NELL-995-Fed3 are 1.76\% and 2.12\%, respectively. The remaining three metrics also show the similar growth trend, overall. This further substantiates the efficacy of utilizing client-wise relation graph for the FKGE task. Although there is more improvement space for \ms and \mpl on FB15k-237-Fed10, FB15k-237-Fed5 and FB15k-237-Fed3 when using ComplEx as KGE method, both \ms and \mpl achieve comparable results with Collective on NELL-995-Fed3 overall.

\begin{table}[h]
\caption{The MRR values on four datasets of \mpl and its variants: PFedEG+\_Reg and PFedEG+\_Init. The \textbf{bold} figures denote the best results.}\label{variants}
\setlength{\tabcolsep}{1mm}{
\begin{tabular}{@{}llllllll@{}}
\toprule%

\multirow{2}{*}{Dataset} &  \multirow{2}{*}{KGE}  && \multicolumn{1}{c}{\mpl} && \multicolumn{1}{c}{PFedEG+\_Reg} && \multicolumn{1}{c}{PFedEG+\_Init}\\

\cmidrule{4-8}
&&&MRR&&MRR && MRR\\

\midrule
\multirow{3}{*}{FB15k-237-Fed10} & TransE&&  $\mathbf{0.4203}$  && 0.4170  && $\mathbf{0.4203}$ \\

                                & RotatE&&  $\mathbf{0.4473}$  && 0.4445  && 0.4307 \\
                                & ComplEx&&  $\mathbf{0.3803}$  && 0.3712  && 0.3545\\
                       
\midrule

\multirow{3}{*}{FB15k-237-Fed5} & TransE&& $\mathbf{0.4273}$   && 0.4271  && 0.4232\\

                                & RotatE&& $\mathbf{0.4495}$   && $\mathbf{0.4495}$  && 0.4335\\
                                & ComplEx&& $\mathbf{0.3961}$   && 0.3854  && 0.3809\\

\midrule

\multirow{3}{*}{FB15k-237-Fed3} & TransE&& $\mathbf{0.4357}$   && 0.4335  && 0.4303\\

                                & RotatE&& $\mathbf{0.4539}$   &&  0.4524 && 0.4435\\
                                & ComplEx&& $\mathbf{0.4086}$   && 0.3942  && 0.3956\\
\midrule
\multirow{3}{*}{NELL-995-Fed3} & TransE&&  $\mathbf{0.3687}$  && 0.3679  && 0.3583\\

                                & RotatE&& $\mathbf{0.4423}$   && 0.4312  && 0.4344\\
                                & ComplEx&& $\mathbf{0.3737}$   && 0.3628  && 0.3591\\    
\botrule
\end{tabular}
}
\end{table}

By comparing the performances of both \ms and \mpl on FB15k-237-Fed10, FB15k-237-Fed5, FB15k-237-Fed3 and NELL-995-Fed3 with various client numbers (10, 5, 3 and 3, correspondingly), we find that both \ms and \mpl are more effective with more number of clients overall. For example, compared with the best results of Single, FedE and FedEC, the best performance of \ms and \mpl, with RotatE as KGE method, achieves relative rise in MRR on the four datasets by 4.73\%, 2.72\% 1.72\% and 1.58\%, respectively.

\subsection{Ablation Study}
\begin{figure}[h]
  \centering
    \begin{minipage}{0.35\linewidth}
		\centering
		\includegraphics[width=0.99\linewidth]{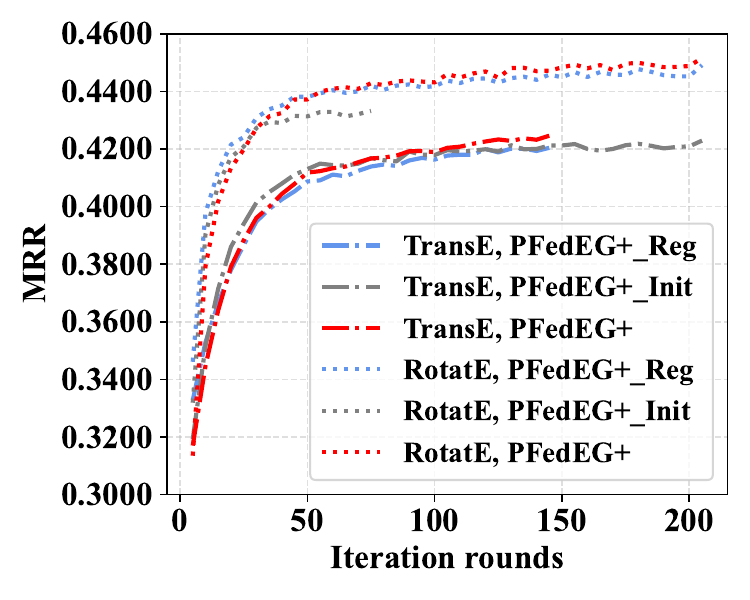}
		\subcaption{\scriptsize{FB15k-237-Fed10}}
		\label{FB15k-237-Fed10}
    \end{minipage}
     \begin{minipage}{0.35\linewidth}
		\centering
		\includegraphics[width=0.99 \linewidth]{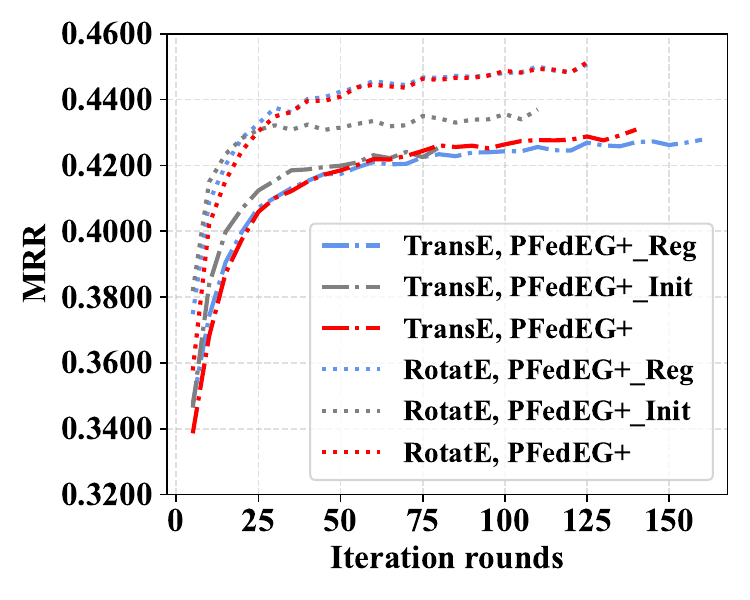}
		\subcaption{\scriptsize{FB15k-237-Fed5}}
		\label{FB15k-237-Fed5}
    \end{minipage}
    \begin{minipage}{0.35\linewidth}
		\centering
		\includegraphics[width=0.99 \linewidth]{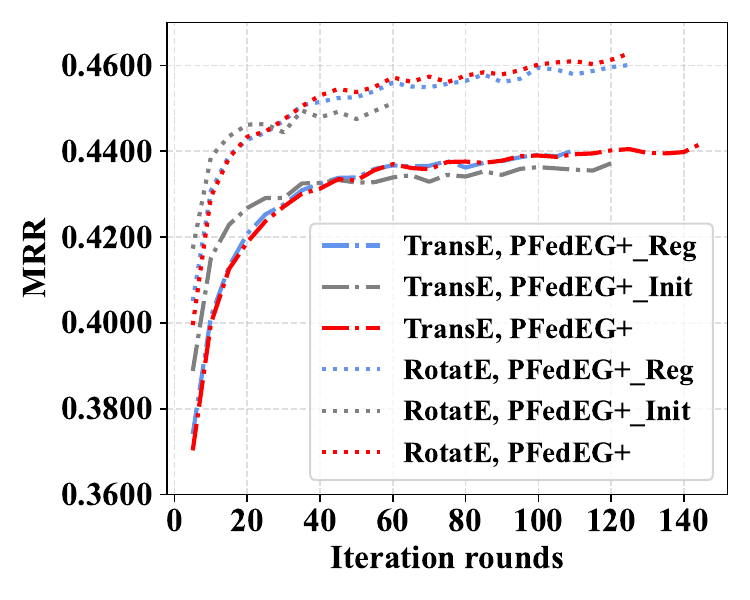}
		\subcaption{\scriptsize{FB15k-237-Fed3}}
		\label{FB15k-237-Fed3}
    \end{minipage}
    \begin{minipage}{0.35\linewidth}
		\centering
		\includegraphics[width=0.99 \linewidth]{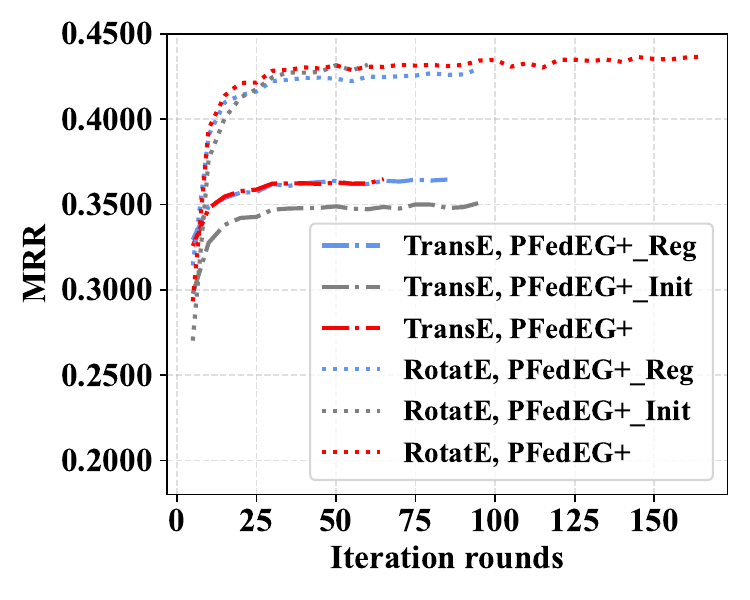}
		\subcaption{\scriptsize{NELL-995-Fed3}}
		\label{NELL-995-Fed3}
    \end{minipage}
  \caption{The Comparison of convergence about \mpl and its variants on validation set of four dataset.}
  \label{variants_fig}
\end{figure}

\begin{figure*}
\vspace{-0.5 em}  
\footnotesize
  \centering
    \begin{minipage}{0.99\linewidth}
		\centering
		\includegraphics[width=0.95\linewidth]{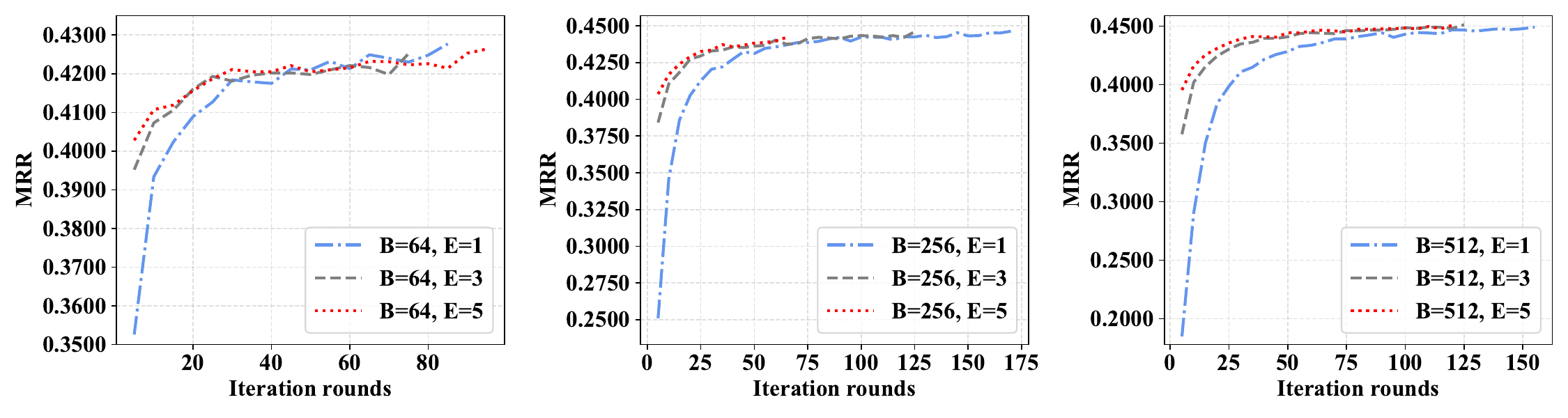}
		\subcaption{\scriptsize{Fixed batch size(B) and various local epochs(E).}}
		\label{FB}
    \end{minipage}
     \begin{minipage}{0.99\linewidth}
		\centering
	\includegraphics[width=0.95\linewidth]{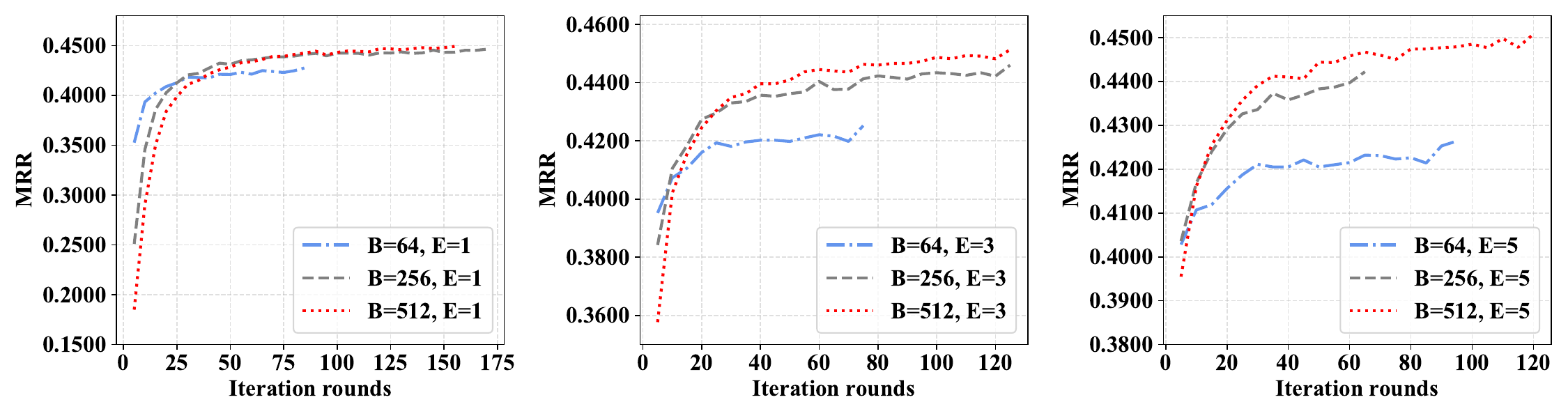}
		\subcaption{\scriptsize{Fixed local epochs(E) and various batch sizes(B).}}
		\label{FE}
    \end{minipage}
  \caption{The MRR value of \mpl on validation set of FB15k-237-Fed5 with fixed local epochs(E) and various batch sizes(B).}
  \label{BE}
\vspace{-1 em}  
\end{figure*}

In this section, we investigates the influence of two key factors on \mpl: (1): the impact of initializing local entity embeddings using personalized supplementary knowledge during each round; (2): the role played by the regularization term governing the interaction between local entity embeddings and personalized supplementary knowledge during client updates. We denote the variants of \mpl only retaining $\textbf{Reg}$ularization term and \textbf{Init}ialization process as $\textbf{PFedEG+\_Reg}$ and $\textbf{PFedEG+\_Init}$, respectively.
In each variant's case, the shared parameters with \mpl remain consistent, and subsequently, link prediction is executed across four distinct datasets as introduced in section \ref{es}.

The results pertaining to MRR are exhibited in Table \ref{variants}. The analysis reveals that the performance of \mpl surpasses that of its variants overall. This performance enhancement is particularly significant when ComplEx is employed as the KGE method. This underscores the efficacy of concurrent utilization of initialization process and regularization term about personalized supplementary knowledge during client-specific embedding learning. 
Besides, it is found that PFedEG+\_Reg outperforms PFedEG+\_Init in most cases, which suggests the regularization term usually can leverage information in personalized supplementary knowledge better. We leave it as future work to study more effective approaches to leverage personalized supplementary knowledge to further improve accuracy of FKGE.

Furthermore, a convergence analysis is conducted on \mpl and its variants, the outcomes of which are depicted in Figure \ref{variants_fig}. For the KGE method ComplEx, both \mpl and its variants exhibit swift convergence, typically requiring a limited number of iterations (generally fewer than 30) to complete the training process. Consequently, we exclude it and only conduct experiment with TransE and RotatE. From Figure \ref{variants_fig}, it is found that PFedEG+\_Init witnesses the lowest convergence rounds in most cases, particularly notable when RotatE is employed as the KGE method for datasets FB15k-237-Fed10, FB15k-237-Fed3, and NELL-995-Fed3. Besides, the convergence rounds required by \mpl and PFedEG+\_Reg are generally comparable, excepting that RotatE is used as the KGE method on the NELL-995-Fed3 dataset.

\subsection{Model Analysis about Various Parameters}

In this part, we investigate how parameter variations influence our proposed model \mpl.

We first investigate the impact of batch size and local epochs on the performance of \mpl on FB15k237-Fed5, with RotatE as the KGE method. Three different local epoch ($\mathbf{1, 3, 5}$), are chosen, along with three different batch sizes ($\mathbf{64, 256, 512}$). The MRR values of \mpl on validation set are presented in Figure \ref{FB} and Figure \ref{FE}. 

From Figure \ref{FB}, we can find that when batch size is fixed, different local epochs hardly affects the performance of \mpl for FKGE. Besides, there is no obvious correlation between the training rounds of \mpl and the local epoch. For example, when batch size is 64 and epoch size is set to the largest, namely 5, \mpl requires the most iteration round while when batch size is 256 (512) and epoch size is set to the largest, \mpl requires the least iteration round. Analysis of the Figure \ref{FE} reveals that when local epoch is fixed, the performance of \mpl for FKGE increases as the increase of batch size. However, \mpl usually converges slower and needs more rounds to conduct training as the increase of batch size. Hence, the selection of batch size and local epoches should consider both expected performance and time consumption.

\begin{figure}
\vspace{-0.5 em}  
\centerline{\includegraphics[height=1.5 in]{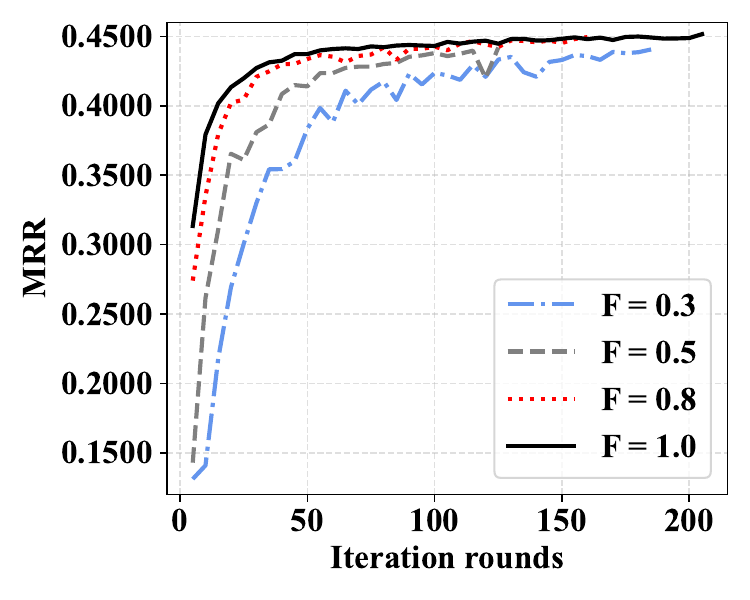}}
\caption{Results with different client selection fraction $\mathbf{F}$. } \label{f10_frac} 
\vspace{-1.5 em} 
\end{figure}

Besides, we investigate the impact of multi-client parallelism on \mpl. Using RotatE as the KGE method, we performed FKGE on FB15k-237-Fed10 with varying client selection fraction ($\mathbf{F = 0.3, 0.5, 0.8, 1}$). The results on the validation dataset are presented in Figure \ref{f10_frac}. The horizontal and vertical coordinates represent the iteration rounds and MRR values in current round respectively. Different curves represent different client selection fractions ($\mathbf{F}$). It is found that \mpl with higher $\mathbf{F}$ usually have higher FKGE performance. Besides, there is no obvious correlation between the training rounds of \mpl and $\mathbf{F}$. For example, when $\mathbf{F}$ is 0.3, the required iteration round is close to the case where $\mathbf{F}$ is 1.0 while is much more than the case where $\mathbf{F}$ is 0.5. However, the consumption of computational resources increases with the increase of $\mathbf{F}$. Therefore, the selection of client selection fraction should take the above aspects into consideration.
\section{CONCLUSIONS}

We proposed \mn, a novel federated knowledge graph embedding method. Compared with existing methods, it introduces the client-wise relation graph to form personalized supplementary knowledge for each client rather than a copy of global knowledge shared by all clients, which can better leverage useful information from other KGs to improve quality of learned embeddings. Moreover, we explain the effect of semantic disparities of FKG and propose to conduct personalized embedding learning for individual KG in FKG using personalized supplementary knowledge and local triples to further improve performance of FKGE, rather than learn a copy of global consensus entity embeddings for all clients as previous methods did. Finally, we conducted extensive experiments on four datasets and the experiment results show the effectiveness of our method. Besides, despite the simplicity of relation weights among clients in this paper, it shows great potential to improve the performance of FKGE. In the future, we will explore more effective methods to learn the relation weights among clients for further improving the performance of FKGE.

\bibliography{sn-bibliography}

\end{document}